\documentclass{article}
\usepackage{chemfig}
\usepackage{mathtools}
\usepackage[version=4]{mhchem}
\usepackage{chemfig}
\usepackage{amsfonts}
\usepackage{amsmath}
\usepackage{amssymb}
\usepackage{graphicx}
\usepackage{booktabs}
\providecommand{\U}[1]{\protect\rule{.1in}{.1in}}

\date{}
\begin{document}

\title{Modelling the Spreading of the SARS-CoV-2 in Presence of the Lockdown and Quarantine Measures by a "Kinetic-Type Reactions" Approach}
\author{SONNINO Giorgio, PEETERS Philippe, and NARDONE Pasquale\\
Universit{\'e} Libre de Bruxelles (ULB), Facult{\'e} de Sciences\\
Bvd du Triomphe, Campus Plaine CP 231, 1050 Brussels, Belgium}
\maketitle
\noindent Emails: gsonnino@ulb.ac.be, peeters.philippe@gmail.com, pnardon@ulb.ac.be
\begin{abstract}
We propose a realistic model for the evolution of the COVID-19 pandemic subject to the lockdown and quarantine measures, which takes into account the time-delay for recovery or death processes. The dynamic equations for the entire process are derived by adopting a \textit{kinetic-type reactions} approach. More specifically, the lockdown and the quarantine measures are modelled by some kind of inhibitor reactions where susceptible and infected individuals can be \textit{trapped} into inactive states. The dynamics for the recovered people is obtained by accounting people who are only traced back to hospitalised infected people. To get the evolution equation we take inspiration from the Michaelis- Menten’s enzyme-substrate reaction model (the so-called \textit{MM reaction}) where the \textit{enzyme} is associated to the \textit{available hospital beds}, the \textit{substrate} to the \textit{infected people}, and the \textit{product} to the \textit{recovered people}, respectively. In other words, everything happens as if the hospitals beds act as a \textit{catalyzer} in the hospital recovery process. Of course, in our case the reverse \textit{MM reactions} has no sense in our case and, consequently, the kinetic constant is equal to zero. Finally, the O.D.E.s for people tested positive to COVID-19 is simply modelled by the following kinetic scheme $S+I\Rightarrow 2I$ with $I\Rightarrow R$ or $I\Rightarrow D$, with $S$, $I$, $R$, and $D$ denoting the compartments Susceptible, Infected, Recovered, and Deceased people, respectively. The resulting \textit{kinetic-type equations} provide the O.D.E.s, for elementary \textit{reaction steps}, describing the number of the infected people, the total number of the recovered people previously hospitalised, subject to the lockdown and the quarantine measure, and the total number of deaths. The model foresees also the second wave of Infection by Coronavirus. The tests carried out on real data for Belgium, France and Germany confirmed the correctness of our model.
\end{abstract}
\section{\bf Introduction}\label{I}
Coronavirus disease 2019 (COVID-19) is caused by a new Coronavirus (SARS-CoV-2) that has spread rapidly around the world. Most infected people have no symptoms or suffer from mild, flu-like symptoms, but some become seriously ill and can die. In recent weeks coronavirus has had too many opportunities to spread again. After successfully tamping down the first surge of infection and death, Europe is now in the middle of a second coronavirus wave as it moves into winter \cite{cacciapaglia}, \cite{baley}, \cite{sonia}, \cite{vynnycky}, \cite{gleick}, \cite{coullet}. Even though several vaccines for COVID-19 are actually been produced other ways of slowing its spread have to continue to be explored. One way of controlling the disease are the lockdown and the quarantine measures. The lockdown measures are emergency measures or conditions imposed by governmental authorities, as during the outbreak of an epidemic disease, that intervene in situations where the risk of transmitting the virus is greatest. Under these measures, people are required to stay in their homes and to limit travel movements and opportunities for individuals to come into contact with each other such as dining out or attending large gatherings. The lockdown measures are more effective when combined with other measures such as the quarantine. Quarantine means separating healthy people from other healthy people, who may have the virus after being in close contact with an infected person, or because they have returned from an area with high infection rates. Similar recommendations include isolation (like quarantine, but for people who tested positive for COVID-19) and physical distancing (people without symptoms keep a distance from each other). Several governments have then decided that stricter lockdown and quarantine measures are needed to bring down the number of infections. In this work we shall propose interventions which are as targeted as possible. Unfortunately, the greater the number of infections, the more sweeping the measures have to be. Tightening the measures will impact on our society and the economy but this step is needed for getting the coronavirus under control.

\noindent The aim of this work is to model the dynamics of the infectious, recovered, and deceased people when population is subject to lockdown and quarantine measures imposed by governments. We shall see that the combined effect of the restrictions measures with the action of the Hospitals and Health Institutes is able to contain and even dampen the spread of the SARS-CoV-2 epidemic. The dynamics of the entire process will be obtained by taking into account the theoretical results recently appeared in literature \cite{sonnino} and \cite{sonnino1} and by adopting a \textit{kinetic-type reactions} approach. In this framework, the dynamics of the Health Institutes is obtained by taking inspiration from the Michaelis- Menten’s enzyme-substrate reaction model (the so-called \textit{MM reaction} \cite{MM1}, \cite{MM2}, and \cite{MM3}) where the \textit{enzyme} is associated to the \textit{available hospital beds}, the \textit{substrate} to the \textit{infected people}, and the \textit{product} to the \textit{recovered people}, respectively. In other words, everything happens as if the hospitals beds act as a \textit{catalyzer} in the hospital recovery process \cite{sonnino3}. In addition, the time-delay for recovery or death processes are duly taken into account. More specifically, in our model, we have the following 10 compartments:

\noindent $S$ = Number of susceptible people. This number concerns individuals not yet infected with the disease at time $t$, but they are susceptible to the disease of the population;

\noindent $I$ = Number of people who have been infected and are capable of spreading the disease to those in the susceptible category;

\noindent $I_h$ = Number of hospitalised infected people;

\noindent $I_Q$ = Number of people in quarantine. This number concerns individuals who may have the virus after being in close contact with an infected person;

\noindent $R$ = Total number of recovered people, meaning specifically individuals having survived the disease and now immune. Those in this category are not able to be infected again or to transmit the infection to others;

\noindent $r_h$ = Total recovered people previously hospitalised;

\noindent $D$ = Total number of people dead people for COVID-19;

\noindent $d_h$ = Total number of people previously hospitalised dead for COVID-19;

\noindent $L$ = Number of inhibitor sites mimicking lockdown measures:

\noindent $Q$ = Number of inhibitor sites mimicking quarantine measures.

\noindent In addition, $N$, defined in Eq.~(\ref{6.4}), denotes the number of total cases.

\noindent The manuscript is organised as follows. In Section~\ref{ODE} we derive the deterministic Ordinary Differential Equations (ODSs) governing the dynamics of the infectious, recovered, and deceased people. The lockdown and quarantine measures are modelled in Subsection~\ref{LQM}. The dynamics of the hospitalised individuals (i.e., the infectious, recovered, and deceased people) can be found in Subsection~\ref{H}. As mentioned above, the corresponding ODEs are obtained by considering the \textit{MM reaction model}. The equations governing the dynamics of the full process and the related \textit{basic reproduction number} are reported in Section~\ref{TODEs} and Section~\ref{BRN}, respectively. It is worth mentioning that our model foresees also the second wave of Infection by Coronavirus. As shown in Section~\ref{SIRD}, in absence of the restrictive measures and by neglecting the role of the Hospitals and the delay in the reactions steps, our model reduces to the classical  \textit{Susceptible-Infectious-Recovered-Deceased-Model} (SIRD-model) \cite{sird}. Finally, Section~\ref{Applications} shows the good agreement between the theoretical predictions with real data for Belgium, France and Germany. The last Section~\ref{C} presents the conclusions and perspectives of this manuscript.

\section{\bf Model for COVID-19 in Presence of the Lockdown and Quarantine Measures}\label{ODE}

\noindent As mentioned in the Introduction, the population is assigned to compartments with labels $S$, $I$, $R$ $D$ etc. The dynamics of these compartments is generally governed by deterministic ODEs, even though stochastic differential equations should be used to describe more realistic situations \cite{sonnino}. In this Section, we shall derive the deterministic ordinary differential equations obeyed by compartments. This task will be carried out by taking into account the theoretical results recently appeared in literature \cite{sonnino1}, \cite{sonnino2} and without neglecting the delay in the reactions steps.

\subsection{Modelling the Susceptible People}

\noindent If a susceptible person encounters an infected person, the susceptible person will be infected as well. So, the scheme simply reads
\begin{equation}\label{S1}
S + I \xrightarrow{\mu} 2I
\end{equation}
\subsection{Modelling the Lockdown and Quarantine Measures}

\noindent The lockdown measures are mainly based on the isolation of the susceptible people, (eventually with the removal of infected people by hospitalisation), but above all on the removal of susceptible people. 
\vskip0.2cm
\noindent {\bf Subsection 2.1. Modelling the Lockdown and Quarantine Measures with Chemical Interpretation}\label{LQM}

\begin{figure}[hbt!]
\hskip 0.5truecm
\includegraphics[width=10cm, height=6cm]{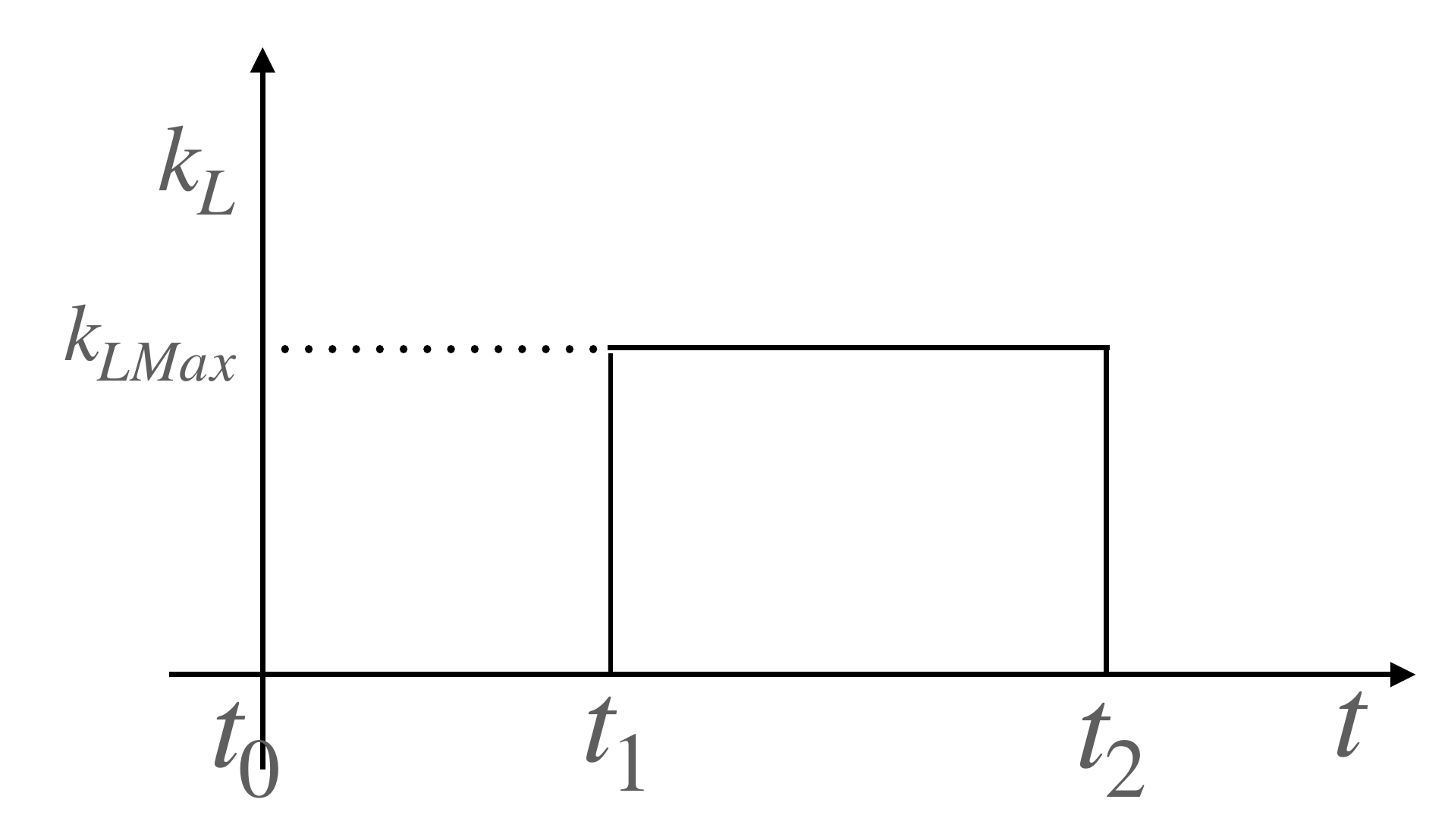}
\caption{
\textit{{\bf Lockdown Efficiency Parameter.} For simplicity, in our model the lockdown efficiency parameter $k_L$ is a \textit{door-step function}. This function is constant, $K_{LMax}\neq 0$,within the range $t_1\leq t\leq t_2$ and zero outside it.} 
}
\label{LEP}
\end{figure}

\noindent It is assumed the lockdown and quarantine measures are modelled by some kind of inhibitor reaction where the susceptible people and the infected can be \textit{trapped} into inactive states $S_L$ and $I_Q$, respectively. Indicating with $L$ and $Q$ the Inhibitor sites mimicking the lockdown and the quarantine measures respectively, we get
\begin{align}\label{LQM1}
&S + L \xrightleftharpoons[k_{LMax}-k_L]{k_L} S_L\\
&I  \xrightarrow{k_Q} I_Q\xRightarrow{k_{QR}, \ t_{QR}} R\nonumber
\end{align}
\noindent In the scheme~(\ref{LQM1}), symbol $\implies$ stands for a \textit{delayed reaction} just like \textit{enzyme degradation processes} for instance. Here, $L_{max}=S_L+L$ hence, if $L\simeq L_{Max}$, an almost perfect lockdown measures would totally inhibit virus propagation by inhibiting all the susceptible people $S$ and the infected people $I$. A not so perfect lockdown measures would leave a fraction of $I$ free to spread the virus. The number of inhibitor sites maybe a fraction of the number of the infected people. Fig.~\ref{LEP}. shows the behaviour of the lockdown efficiency parameter adopted in our model. For simplicity, we have chosen a parameter which is constant $k_{LMax}\neq0$ inside the time-interval $t_1\leq t\leq t_2$ and vanishes outside it. The \textit{inverse Lockdown efficiency parameter} is $k^{-1}_L=k_{LMax}-k_L$, which is equal to $k_{LMax}$ outside the door and vanishes inside the the interval $t_1\leq t\leq t_2$.
%%%%%%%%%%%%%%%%%%%%%%%%%%%%%%%%%%%%%%%%%%
%%%%%%%%%%%%%%%%%%%%%%%%%%%%%%%%%%%%%%%%%%
Finally, from Schemes~(\ref{S1}) and (\ref{LQM1}), we get the O.D.E.s for $S$, $L$, $Q$, and $I_Q$:
\begin{align}\label{LQM2}
&{\dot S}=-\mu SI - k_L S(L_{Max}-S_L)+(1-k_L)(L_{Max}-L)\\
&{\dot S}_L=k_L SL-k^{-1}_LS_L\nonumber\\
&{\dot I_Q}=k_Q I-\chi{ I_Q}_{(t-t_R)}\noindent
\end{align}
\noindent with the \textit{dot} above the variables denoting the \textit{time derivative}.
\noindent 

\subsection{ O.D.E. for the Total Recovered People}

\noindent At the first approximation, the O.D.E. for the \textit{total recovered people} $R$ (i.e. the total individuals having survived the disease) is trivially obtained by considering the following \textit{kinetic scheme}:
\begin{align}\label{R1}
&I \xRightarrow{\chi ,\ t_R} R\\
&I_Q \xRightarrow{k_{QR} ,\ t_{QR}} R\nonumber 
\end{align}
\noindent That is, the rate of $R_t$ is approximatively proportional to the number of the infected people $I$ at time $t$ i.e.\footnote{Notice that the first \textit{reaction} in the scheme Eq.~(\ref{R1}) is the dynamic equation for the total recovered people adopted in the SIRD-model \cite{sird}.}.
\begin{equation}\label{R}
{\dot R}=\chi I_{(t-t_R)}+\chi R_{(t-t_R)} 
\end{equation}
\noindent where we have introduced the time-delay $t_R$ (the number of the recovered people at time time $t$ is proportional to the infected people at time $t-t_R$). However, it is useful to clarify the following. In Eqs~(\ref{R1}), $R$ stands for the \textit{total number of the recovered people} (i.e. the number of the recovered people previously hospitalised, plus the number of the asymptomatic people, plus the infected people who have been recovered without being previously hospitalised). The natural question is: \textit{how can we count $R$ and compare this variable with the real data ?}. The current statistics, produced by the Ministries of Health of various Countries, concern the people released from the hospitals. Apart from Luxembourg (where the entire population has been subject to the COVID-19-test), no other Countries are in a condition to provide statistics regarding the total people recovered by COVID-19. Hence, it is our opinion that the equation for $R$, is not useful since it is practically impossible to compare $R$ with the experimental data. We then proceed by adopting approximations and to establish the differential equation whose solution can realistically be subject to experimental verification. More specifically:

\noindent Firstly, we assume that $R$ is given by three contributions:
\begin{equation}\label{R2}
R=r_h+r_{A}+r_{I}
\end{equation}
\noindent with $r_h$, $r_{A}$, and $r_{I}$ denoting the \textit{total number of the recovered people previously hospitalised}, \textit{the total number of asymptomatic people}, and the \textit{total number of people immune to SARS-CoV-2}, respectively. 

\noindent Secondly, we assume that the two contributions $r_{A}$  and $r_{I}$ are negligible i.e. we set $r_A\approx 0$ and $r_{I}\approx 0$ \footnote{We consider that the SARS-CoV-12  has just appeared for the first time. So, we do not consider the asymptomatic people who are immune to the virus without any medical treatment.}. 

\subsection{O.D.E. for the Recovered People in the Hospitals}\label{H}

\noindent Now, let us determine the dynamics for the recovered people in the hospitals. So, we account people who are only traced back to hospitalised infected people. We propose the following model\footnote{Our model is inspired by \text{Michaelis-Menten's enzyme-substrate reaction}. Of course, the reverse \textit{MM reaction} has no sense in our case and, consequently, the \textit{kinetic constant} is equal to zero.}:
\begin{align}\label{H1}
&I + b_h \xrightarrow{k_1} I_h\xRightarrow{k_r, \ t_r} r_h+b_h\\
&\qquad\qquad\ \! I_h \xRightarrow{k_d,\ t_d} d_h+b_h\nonumber
\end{align}
\noindent with $b_h$ denoting the number of available \textit{hospital beds}, $I$ the number of \textit{infected people}, $I_h$ the number of \textit{infected people blocking an hospital bed}, $r_h$ the number of \textit{recovered people previously hospitalised}, and $d_h$ the number of \textit{people deceased in the hospital}. Of course, 
\begin{equation}\label{H2}
I_h+b_h=C_h=const.\qquad {\rm where}\quad{C_h={\rm Total\ hospital's\ capacity}}
\end{equation}
\noindent The dynamic equations for the processes are then:
\begin{align}\label{H3}
&{\dot I}_h=k_1I(C_h-I_h)-k_r{I_h}_{(t-t_r)}-k_d{I_h}_{(t-t_d)}\\
&{\dot r}_h=k_r{I_h}_{(t-t_r)}\nonumber\\
&{\dot d}_h=k_d {I_h}_{(t-t_d)}\nonumber
\end{align}
\noindent where $t_r$ and $t_d$ are the \textit{average recovery time delay} and the \textit{average death time delay}, respectively, and we have taken into account Eq.~(\ref{H2}) i.e., $b_h=C_h-I_h$. In general $t_r\neq t_d\neq 0$. Of course, the variation of $r(t)$ over a period $\Delta t$ is:
\begin{equation}\label{H4}
\Delta {r_h}_t={r_h}_t-{r_h}_{(t-\Delta t)}
\end{equation}

\subsection{O.D.E. for People Tested Positive to COVID-19}

\noindent The number of the infected people may be modelled by the following \textit{kinetic scheme}
\begin{align}\label{I1}
&S + I \xrightarrow{\mu} 2I\\
&I \xRightarrow{\chi ,\ t_R} R\nonumber\\
&I \xRightarrow{\alpha ,\ t_D} D\nonumber\\
&I + b \xrightarrow{k_1} I_h\nonumber\\
&I \xrightarrow{k_Q} I_Q\nonumber
\end{align}
\noindent The scheme~(\ref{I1}) stems from the following considerations
\begin{description}
\item{{\bf a)}} If a susceptible person encounters an infected person, the susceptible person will be infected ;
\item{{\bf b)}} The infected people can either survive and, therefore, be recovered after an average time-delay $t_R$, or die after an average time-delay $t_D$;
\item{{\bf c)}} The schemes~(\ref{LQM1}) and (\ref{H1}), respectively, have been taken into account.
\end{description}
\noindent The differential equation for the infected people is reads then
\begin{equation}\label{I2}
{\dot I}=\mu SI-k_Q IQ-k_1I(C_h-I_h)-\chi I_{(t-t_R)}-\alpha  I_{(t-t_D)}
\end{equation}

\subsection{ O.D.E. for Deaths}

\noindent In this model, we assume that the rate of death is proportional to the infected people, according to the scheme~(\ref{I1}). By also taking into account the scheme~(\ref{LQM1}), we get 
\begin{equation}\label{D1}
I \xRightarrow{\alpha ,\ t_D} D
\end{equation}
\noindent and the corresponding O.D.E. for deaths reads
\begin{equation}\label{D2}
{\dot D}=\alpha I_{(t-t_D)}
\end{equation}

\section{Set of O.D.E.s for the Spread of SARS-CoV-2 when the Lockdown and the Quarantine Measures are Adopted}\label{TODEs}

\noindent By collecting the above O.D.E.s, we get the full system of differential equations governing the dynamics of the number of the infected people, the total number of the recovered people previously hospitalised and the total number of deceased peopled, when the lockdown and the quarantine measures are adopted
\begin{align}\label{6.1}
&{\dot S}=-\mu SI - k_L S(L_{Max}-S_L)+k^{-1}_LS_L\qquad{\rm with}\quad k^{-1}_L=k_{Max}-k_L\\
&{\dot S}_L=-k_L S(L_{Max}-S_L)+k^{-1}_LS_L\nonumber\\
&{\dot I}=\mu SI-k_Q I-k_1I(C_h-I_h)-\chi I_{(t-t_R)}-\alpha  I_{(t-t_D)}\nonumber\\
&{\dot I}_{h}=k_1I(C_h-I_h)-k_r{I_h}_{(t-t_r)}-k_d{I_h}_{(t-t_d)}\nonumber\\
&{\dot I}_{Q}={k_{Q}}I_t-\chi {I_Q}_{(t-t_{R})}\nonumber\\
&{\dot r}_h=k_r{I_h}_{(t-t_r)}\nonumber\\
&{\dot R}=\chi I_{(t-t_R)}+\chi {I_Q}_{(t-t_{R})}\nonumber\\
&{\dot d}_h=k_d {I_h}_{(t-t_d)}\nonumber\\
&{\dot D}=\alpha I_{(t-t_D)}\nonumber
\end{align}
\noindent From Eqs~(\ref{6.1}) we get 
\begin{equation}\label{6.2}
S+S_L+I+I_Q+I_h+R+r_h+D+d_h=const.
\end{equation}
\noindent or, by taking into account that $S+S_L=S_{Tot.}$, $R+r_h=R_{Tot.}$, $D+d_h=D_{Tot.}$, and $I+I_Q+I_h=I_{Tot.}$ we get
\begin{equation}\label{6.3}
S_{Tot.}+I_{Tot.}+R_{Tot.}+D_{Tot.}=const.
\end{equation}
\noindent The number of total cases $N$ is defined as
\begin{equation}\label{6.4}
N=I_{Tot.}+r_h+D_{Tot.}
\end{equation}

\section{The Basic Reproduction Number}\label{BRN}

\noindent We note that, in absence of the lockdown and the quarantine measures, the dynamics of the infectious class depends on the following ratio:
\begin{equation}\label{7.1}
R_0= \frac{\mu}{\chi+\alpha} \frac{S}{N_{Tot.}}
\end{equation}
\noindent with $N_{Tot.}$ denoting the \textit{Total Population}. $R_0$ is the \textit{basic reproduction number}. This parameter provides the expected number of new infections from a single infection in a population by assuming that all subjects are susceptible \cite{baley}, \cite{sonia}. The epidemic only starts if $R_0$ is greater than $1$, otherwise the spread of the disease stops right from the start.

\section{Comparison with the SIRD model}\label{SIRD}

\noindent The \textit{Susceptible-Infectious-Recovered-Deceased-Model} (SIRD-model) is one of the simplest compartmental models, and many models may be derived from this basic form. According to the SIRD model, the dynamic equations governing the above compartments read \cite{sird}
\begin{align}\label{8.1}
&{\dot S}=-\mu S I\\
&{\dot I}=\mu S I-\chi I-\alpha I\nonumber \\
&{\dot R}=\chi I\nonumber\\
&{\dot D}=\alpha I\nonumber
\end{align}
\noindent It is easily checked that Eqs~(\ref{6.1}) reduce to Eqs~(\ref{8.1}) by adopting some assumptions. In particular:

\noindent {\bf 1)} The system is not subject to the lockdown and quarantine measures;

\noindent {\bf 2)} The average times-delay may be neglected;

\noindent {\bf 3)} Hospitals do not enter in the dynamics.

\noindent Under these assumptions, Eqs~(\ref{6.1}) reduce to the SIRD equations:
\begin{align}\label{8.2}
&{\dot S}\simeq -\mu S I\\
&{\dot I}\simeq\mu S I-\chi I-\alpha I\nonumber\\
&{\dot R}=\chi I\nonumber\\
&{\dot D}= \alpha I\nonumber
\end{align}

\section{Application of the Model and Appearance of the Second Wave of SARS-CoV-2 Infection}\label{Applications}

\begin{figure}[h]
\begin{center}
 \includegraphics[width=0.8\textwidth]{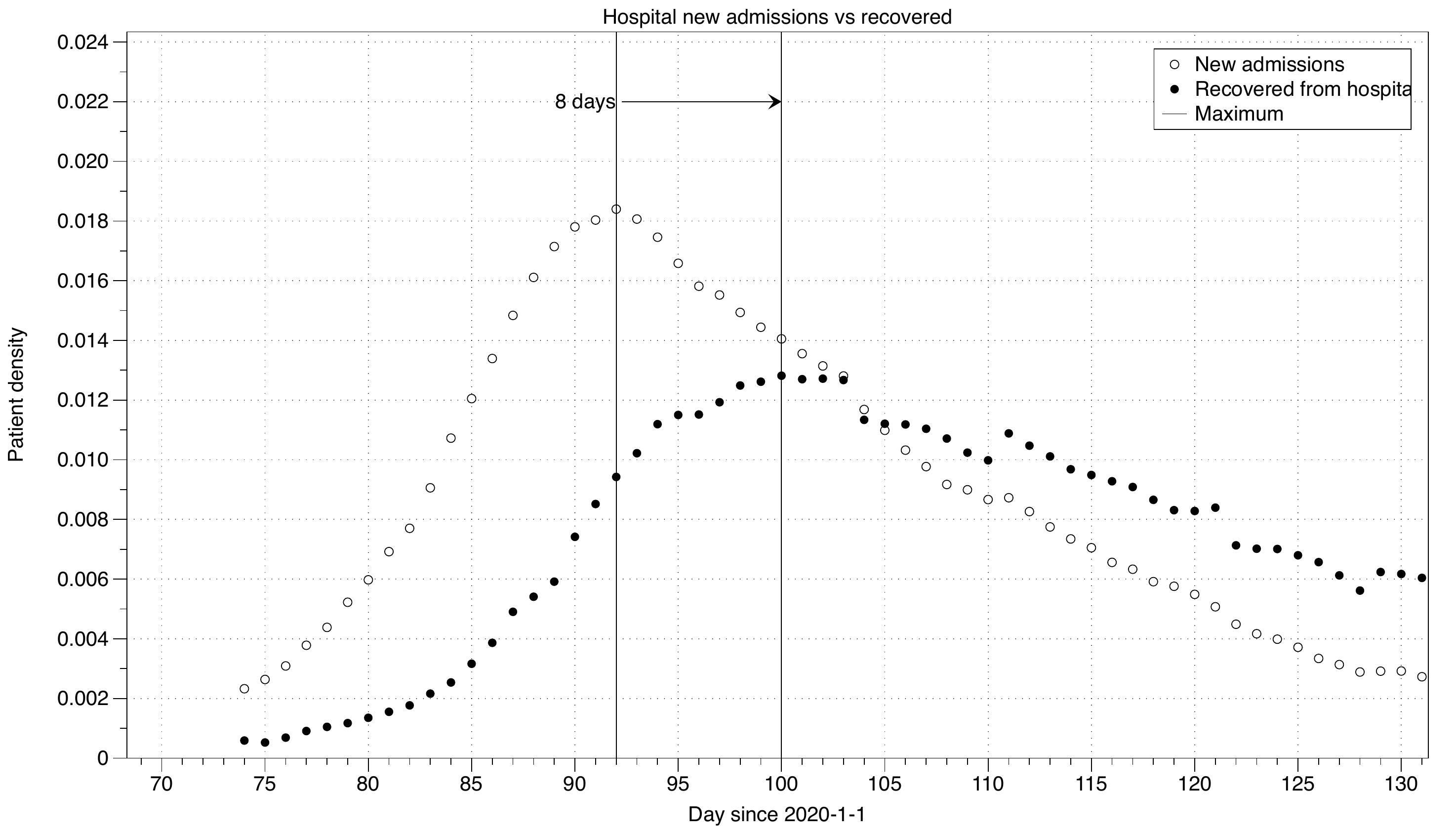}
\caption{\textit{Estimation of the time-delay. The time-delays have been estimated by considering the \textit{time-interval} between the peak of the new admission and the peak of the recovered people from hospitals. This figure corresponds to the Belgian case.}}
\label{delay}
\end{center}
\end{figure}

\noindent Let us now apply our model to the case of a small Country, Belgium, and to other two big Countries, France and Germany. Real data are provided by the various National Health agencies (Belgium - \textit{Sciensano} \cite{dataBE}; France - \textit{Sant{\'e} Publique France} \cite{dataFR}; Germany -\textit{Robert Koch Institut. Country data from Worldbank.org} \cite{dataDE}) and compiled, among others, by European Centre for Disease Prevention and Control (ECDC). It should be noted that this measures does not generally provide the true new cases rate but reflect the overall trend since most of the infected will not be tested \cite{ourworldindata}. It should also be specified that real data provided by ECDC refer to the \textit{new cases per day}, which we denote by $\Delta I_{new}(t)$. By definition, $\Delta I_{new}(t)$ corresponds to the new infected people generated from step $I+S\xrightarrow{\mu} 2I$ solely during 1 day, and \textit{not} to the compartment $I$. Hence, the ECDC data have to be confronted vs the theoretical predictions provided by the solutions for $S(t)$ and $S_{L}(t)$ of our model, according to the relation $\Delta  I_{new}(t) = -\Delta S(t) -\Delta S_{L}(t)$. The values of the parameters used to perform these comparisons are shown in Table~\ref{table}. 
\begin{table}[htp]
\caption{List of the Parameters}
\begin{center}
\begin{tabular}{|l|c|c|c|}
\hline
Parameters & Belgium & France & Germany\\
\hline
 Density [$km^{-2}$] & 377 & 119 & 240 \\
        Surface [$km^{2}$]& 30530 & 547557 & 348560 \\
        $\mu$ [$d^{-1} km^{2}$]& 0.00072 & 0.002 & 0.00093 \\
        $\mu$ after $L_{1}$ &  0.000288 & 0.00087 & 0.000387\\
        $\chi$ [$d^{-1}$]& 0.062 & 0.062 & 0.0608 \\ 
        $\alpha$ [$d^{-1}$]& 0.05 $\chi$  & 0.05 $\chi$  & 0.02 $\chi$ \\
        $k_{L} $  [$d^{-1}$]& 0.07 & 0.06 & 0.06 \\
        $k_Q$ [$d^{-1}$]& 0.02 & 0.01 & 0.01 \\
        $L_m$ [$km^{-2}$]  & 377.0 & 119 & 240 \\
        $k_1$ [$d^{-1} km^{2}$] & 0.01 & 0.01 & 0.01 \\
$k_d + k_r$ [$d^{-1}$] & 0.2 & 0.2 & 0.21 \\
        $\frac{k_d}{k_r} $ & 0.5  & 0.5 & 0.1 \\
        $t_r $ [$d$]& 7  & 7 & 7 \\
        $t_d$ [$d$]& 7 & 7 & 7\\
        $t_R$ [$d$]& 8 & 8 & 8 \\
        $t_D$ [$d$]& 8 & 8 & 8 \\
        $C$ [$km^{-2}$]  & 0.0655 &  0.0091 & 0.023 \\
        $I(60)$ [$km^{-2}$]  & 0.0023 & 0.0018 & 0.0014 \\
        Start $L_{1}$ [$d$]& 77 & 71 & 76 \\
        End $L_{1}$ [$d$]& 124 & 131 & 125 \\
        Start $L_{2}$ [$d$]& 306 & 303 & 306 \\
        \hline
\end{tabular}
\end{center}
\label{table}
\end{table}
\noindent Initial $\mu$ en $k_1$ values have been estimated (fitted) from the measurements using the short period at the start of the pandemic using simple solution valid during that period. $I(60)$ (from March 1, 2020). Hospital capacity is evaluated from the different Countries published capacity. However, we are aware that the interpretation may vary from one Country to another. During the first lockdown, Countries have taken various actions to limit Coronavirus spreading (social distancing, wearing masks, reducing high density hotspots etc.). In order to include these measures in a simple way, we assumed that the net effect is to reduce the actual infection kinetic rate $\mu$ by some constant factor. This is given in the table as $\mu$ after $L_1$. Note that the transition occurs instantaneously in our model hence the sharp drop in the total infected at that time. Other parameters are tuned to account for the actual variability of $\Delta I_{new}$ (but not its absolute value) and official number of deaths ($D(t) + d(t)$). The delay for recovery or death processes has been estimated from the measurements of hospitalisation recovery in a Country. For instance, Fig.~\ref{delay} shows the estimation of the recovery time-delay for Belgium: it corresponds to the \textit{time-interval} between the peak of the new admission and the peak of the recovered people from hospitals. Such a procedure has been adopted for estimating the recovery and death time-delays also for France and Germany.
%%%%%%%%%%%%%%%%%%%%%%%%%%%%%%%%%%%%%%%%

%%%%%%%%%%%%%%%%%%%%%%%%%%%%%%%%%%%%%%%%
\noindent $\bullet$ {\bf Belgian Case}.

\begin{figure*}[htb]
  \hfill
  \begin{minipage}[t]{.48\textwidth}
    \centering
    \includegraphics[width=5cm,height=5cm]{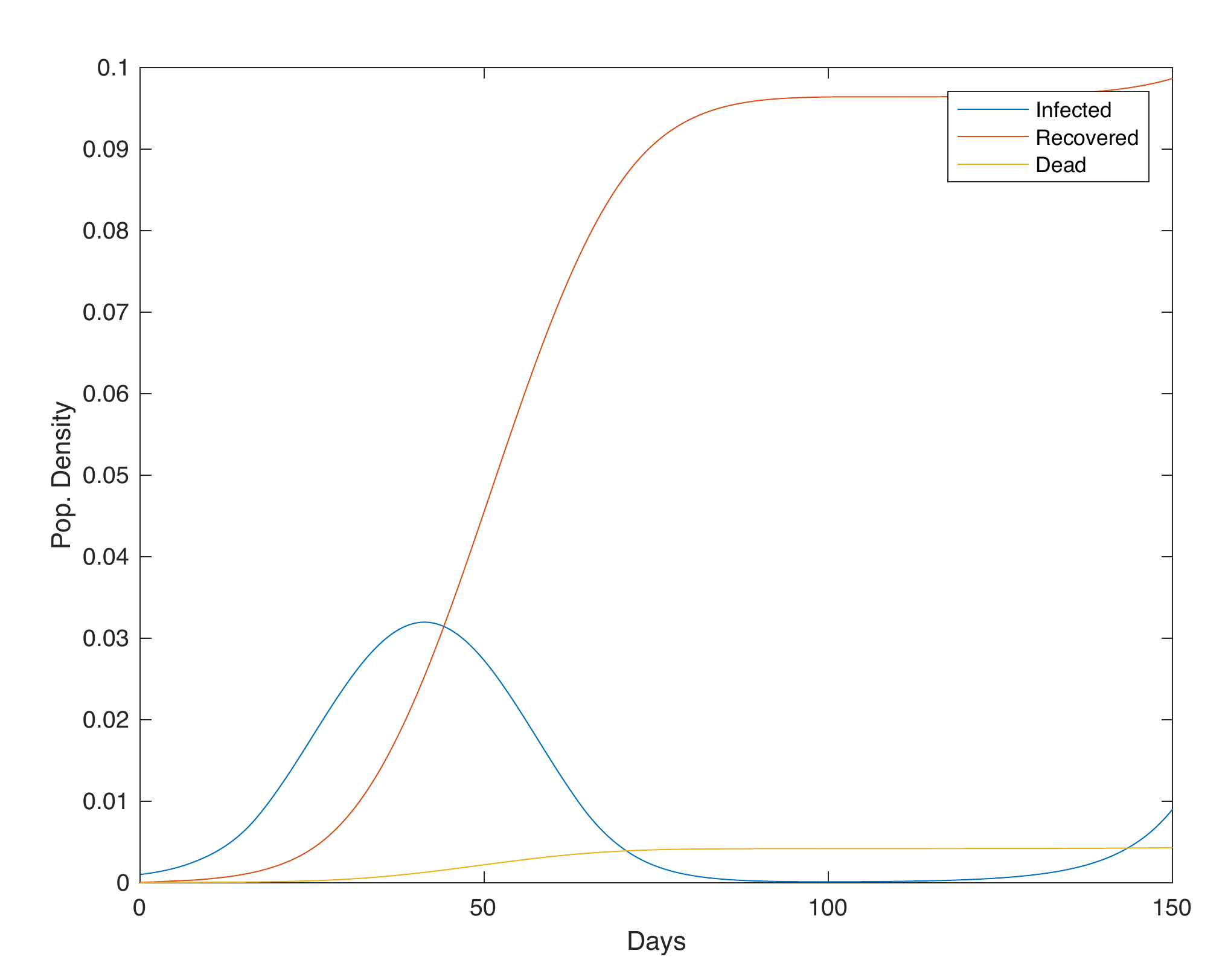}
    \caption{\textit{Theoretical solutions for infectious ($I$), cumulative number of recovered  people ($R$) and deaths ($D$) for Belgium.}}
    \label{BE_IRD}
  \end{minipage}
  \hfill
    \begin{minipage}[t]{.48\textwidth}
      \centering
      \includegraphics[width=5cm,height=5cm]{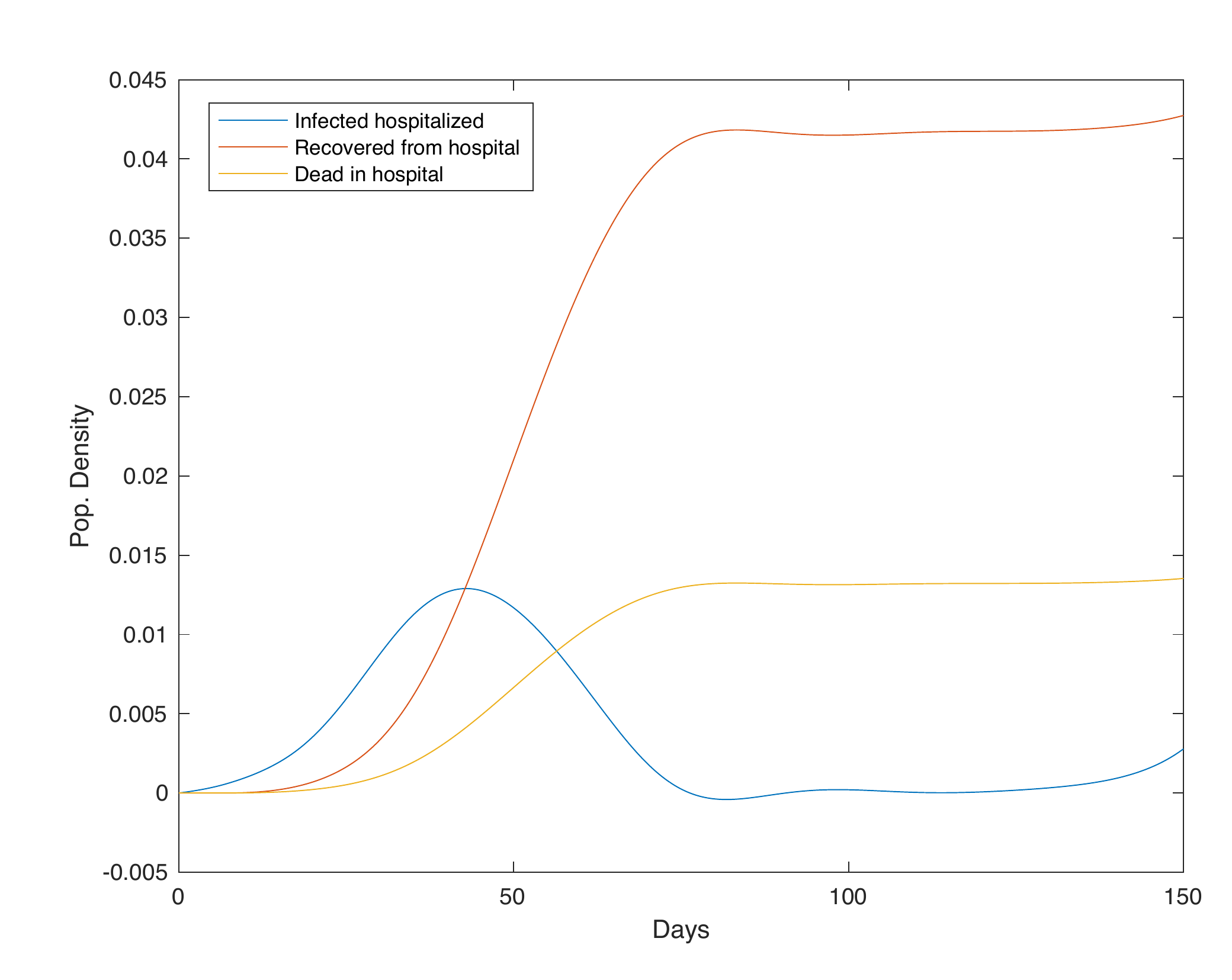}
      \caption{\textit{Theoretical solutions for hospitalised infectious ($I_h$), total recovered ($r_h$) and total deceased ($d_h$) people, previously hospitalised, for Belgium.}}
      \label{BE_IRD_h}
    \end{minipage}
  \hfill
\end{figure*}

\begin{figure*}[htb]
  \hfill
  \begin{minipage}[t]{.48\textwidth}
    \centering
    \includegraphics[width=5cm,height=5cm]{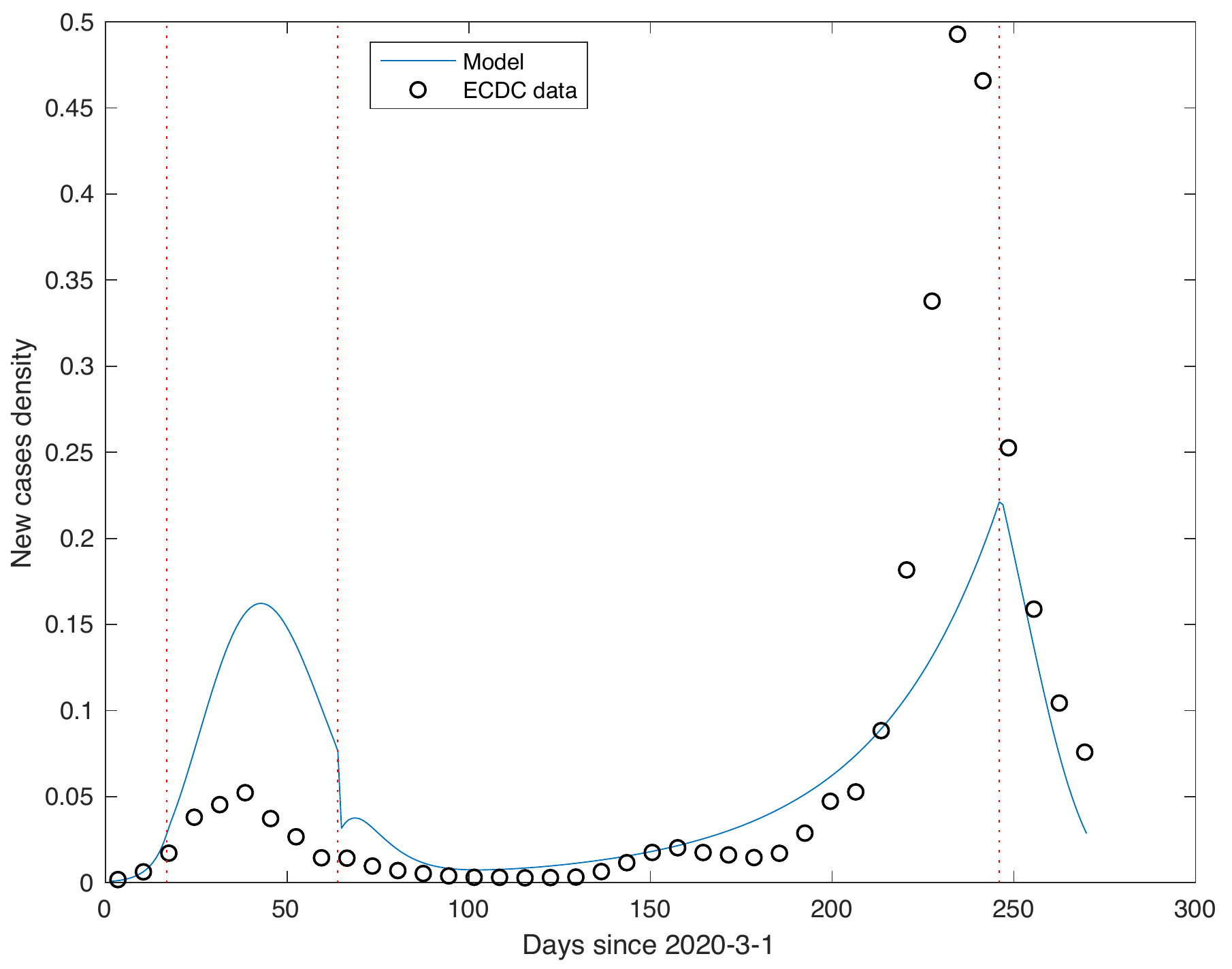}
    \caption{\textit{Comparison between the theoretical prediction for $\Delta I_{New}$ with real data provided by the data base \textit{Sciensano}, for Belgium.}}
    \label{I_new_BE}
  \end{minipage}
  \hfill
    \begin{minipage}[t]{.48\textwidth}
      \centering
      \includegraphics[width=5cm,height=5cm]{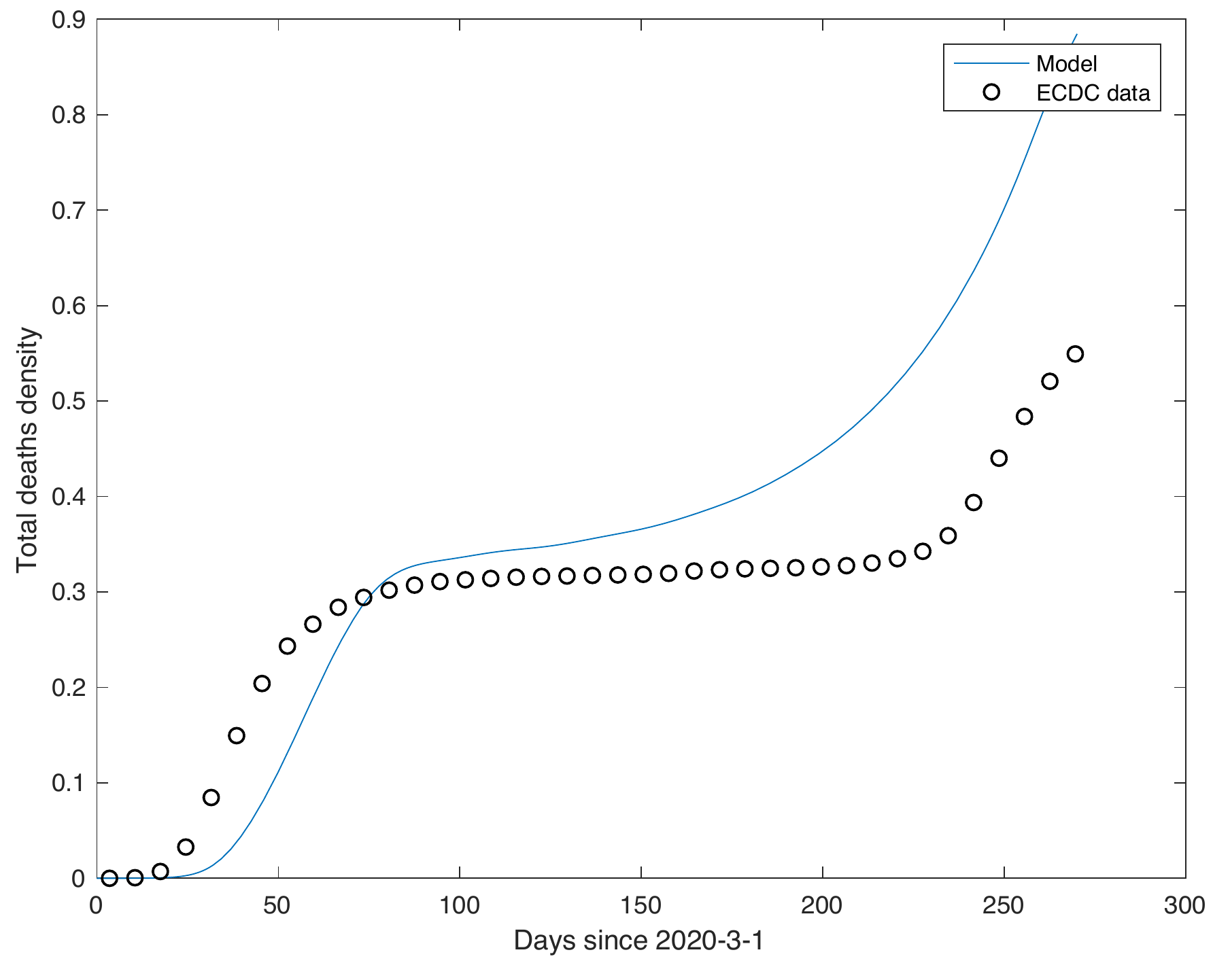}
      \caption{\textit{Comparison between the theoretical solution of our model for Deaths with real data provided by the database \textit{Sciensano}, for Belgium.}}
      \label{D_BE}
    \end{minipage}
  \hfill
\end{figure*}

\noindent Figs~(\ref{BE_IRD}) refer to the Belgian case. In particular, Fig~(\ref{BE_IRD}) shows the solutions of our model for the infectious ($I$), total recovered ($R$) and total deceased ($D$) people. Fig.~(\ref{BE_IRD_h}) illustrates the theoretical solutions for hospitalised infectious ($I_h$), the total recovered ($r_h$) and total deceased ($d_h$) people previously hospitalised. 
%%%%%%%%%%%%%%%%%%%%%%%%%%%%%%%%%%%%%%%%
%%%%%%%%%%%%%%%%%%%%%%%%%%%%%%%%%%%%%%%%
\noindent Figs~(\ref{I_new_BE}) and (\ref{D_BE}) shows the comparison between the theoretical predictions for $\Delta I_{new}(t)$ and deaths and real data for Belgium (according to the database \textit{Sciensano}). Notice in Fig.~\ref{I_new_BE} the prediction of the \textit{second wave of infection by SARS-CoV-2}

%%%%%%%%%%%%%%%%%%%%%%%%%%%%%%%%%%%%%%%%
%%%%%%%%%%%%%%%%%%%%%%%%%%%%%%%%%%%%%%%%
\noindent $\bullet$ {\bf French Case}.

\begin{figure*}[htb]
  \hfill
  \begin{minipage}[t]{.48\textwidth}
    \centering
    \includegraphics[width=5cm,height=5cm]{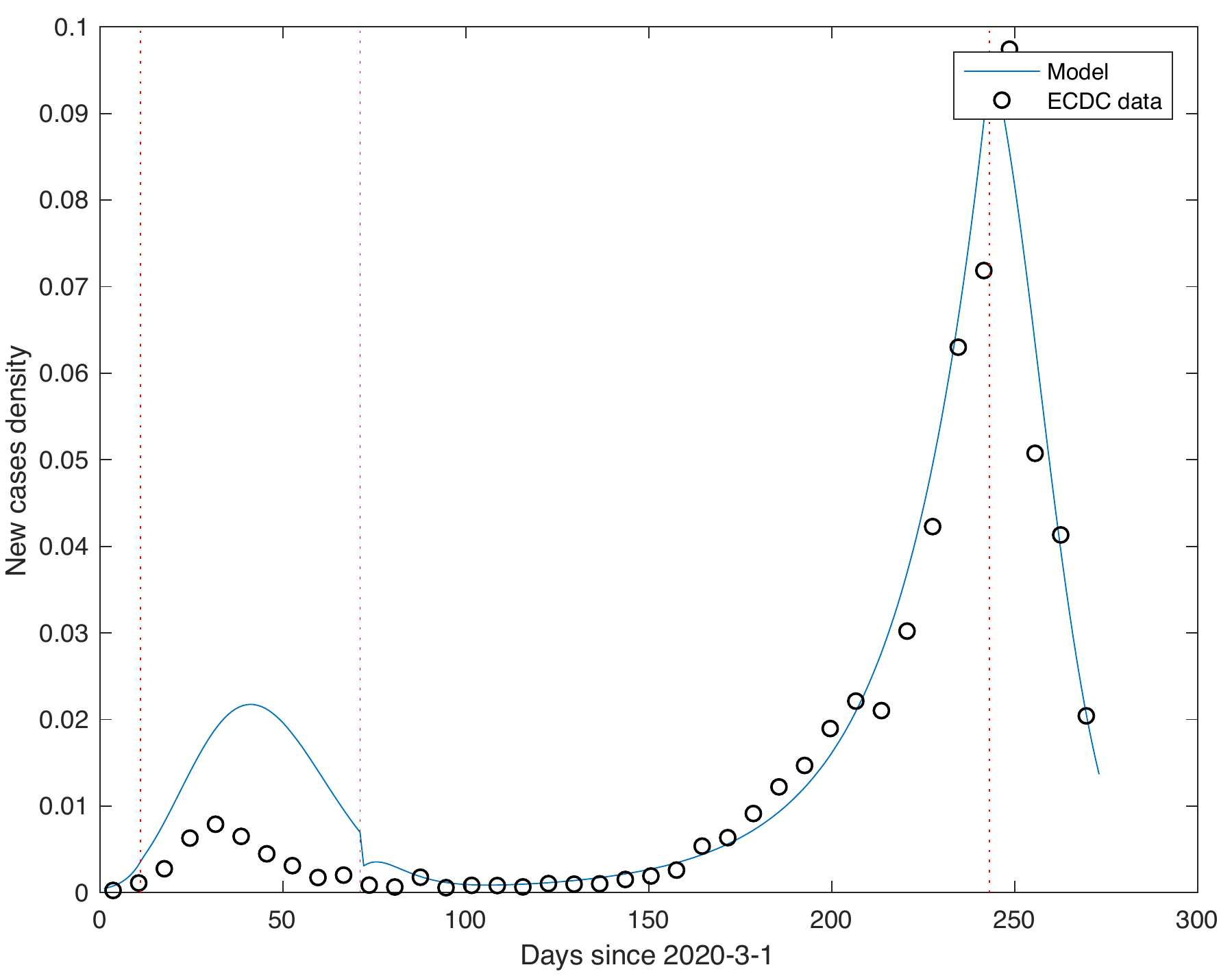}
    \caption{\textit{Comparison between the theoretical prediction for $\Delta I_{New}$ with real data provided by the data base \textit{Sant{\'e} Publique France}, for France.}}
    \label{I_new_FR}
  \end{minipage}
  \hfill
    \begin{minipage}[t]{.48\textwidth}
      \centering
      \includegraphics[width=5cm,height=5cm]{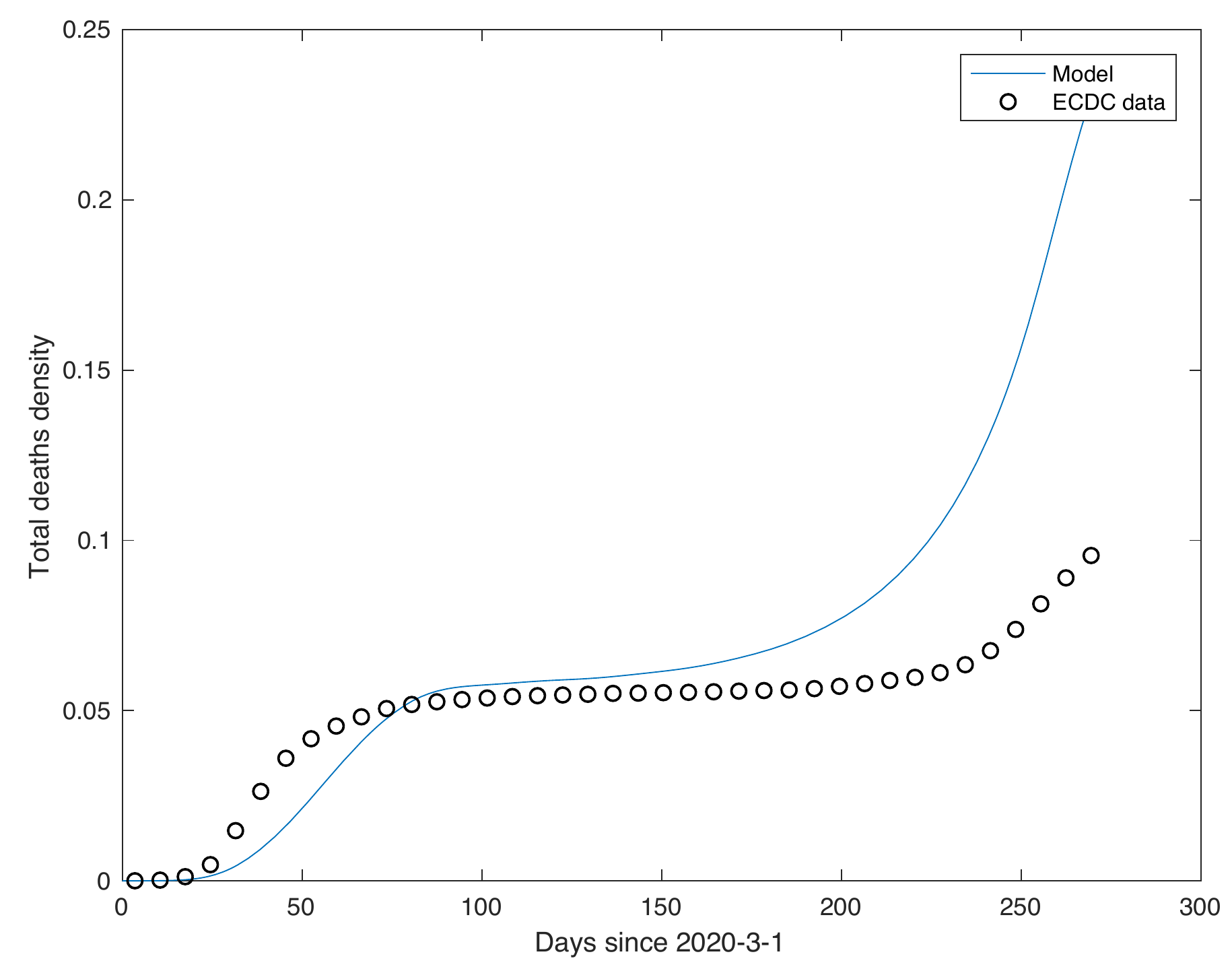}
      \caption{\textit{Comparison between the theoretical solution of our model for Deaths with real data provided by the database \textit{Sant{\'e} Publique France}, for France.}}
      \label{D_FR}
    \end{minipage}
  \hfill
\end{figure*}

\noindent Figs~(\ref{I_new_FR}) and (\ref{D_FR}) shows the comparison between the theoretical predictions for $\Delta I_{new}(t)$ and deaths and real data for Belgium (according to the database \textit{Sant{\'e} Publique France}). Notice in Fig.~\ref{I_new_FR} the prediction of the \textit{second wave of infection by SARS-CoV-2}

%%%%%%%%%%%%%%%%%%%%%%%%%%%%%%%%%%%%%%%%
%%%%%%%%%%%%%%%%%%%%%%%%%%%%%%%%%%%%%%%%
\noindent $\bullet$ {\bf German Case}.

\begin{figure*}[htb]
  \hfill
  \begin{minipage}[t]{.48\textwidth}
    \centering
    \includegraphics[width=5cm,height=5cm]{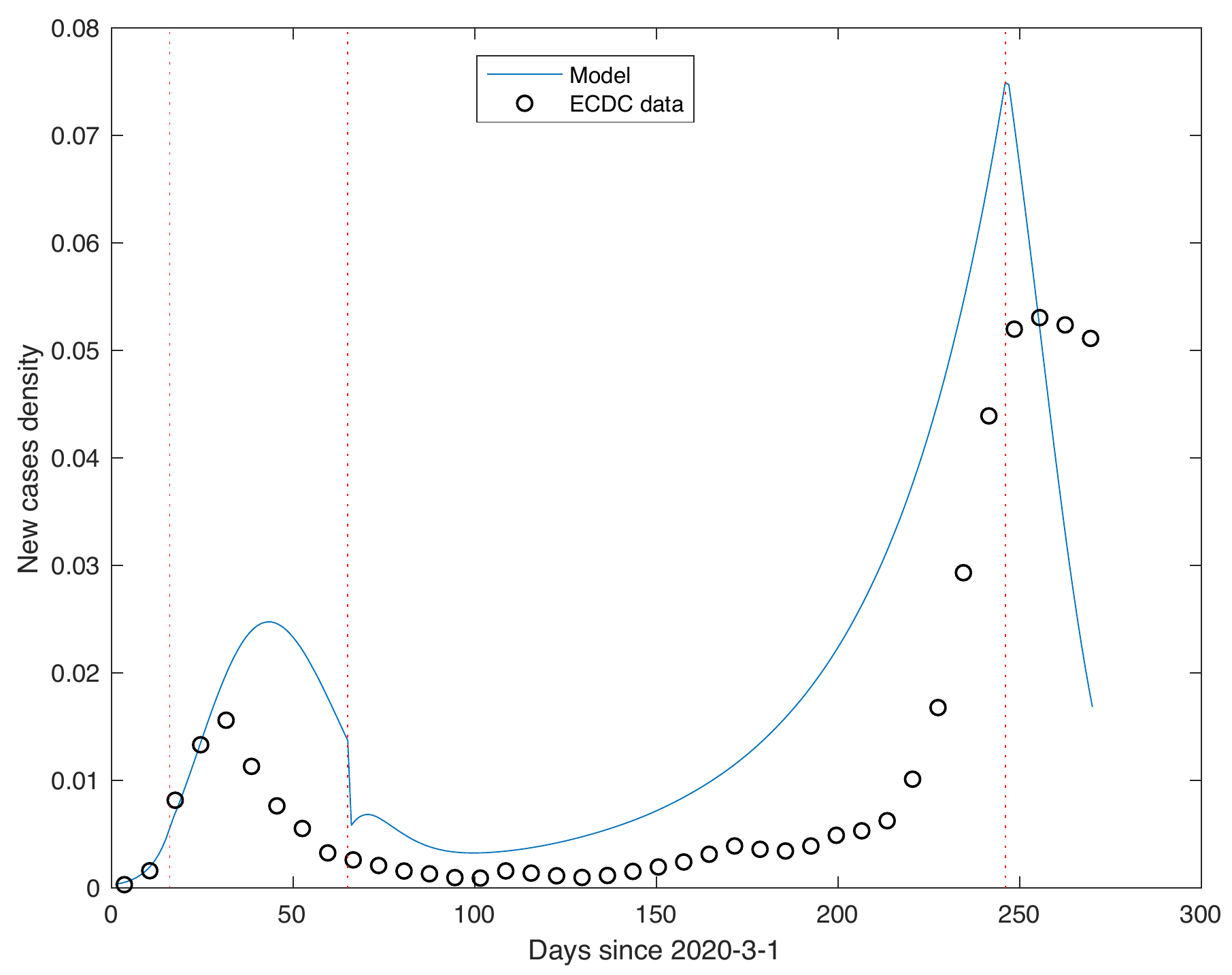}
    \caption{\textit{Comparison between the theoretical prediction for $\Delta I_{New}$ with real data provided by the data base \textit{(Robert Koch Institut. Country data from Worldbank.org}, for Germany.}}
    \label{I_new_DE}
  \end{minipage}
  \hfill
    \begin{minipage}[t]{.48\textwidth}
      \centering
      \includegraphics[width=5cm,height=5cm]{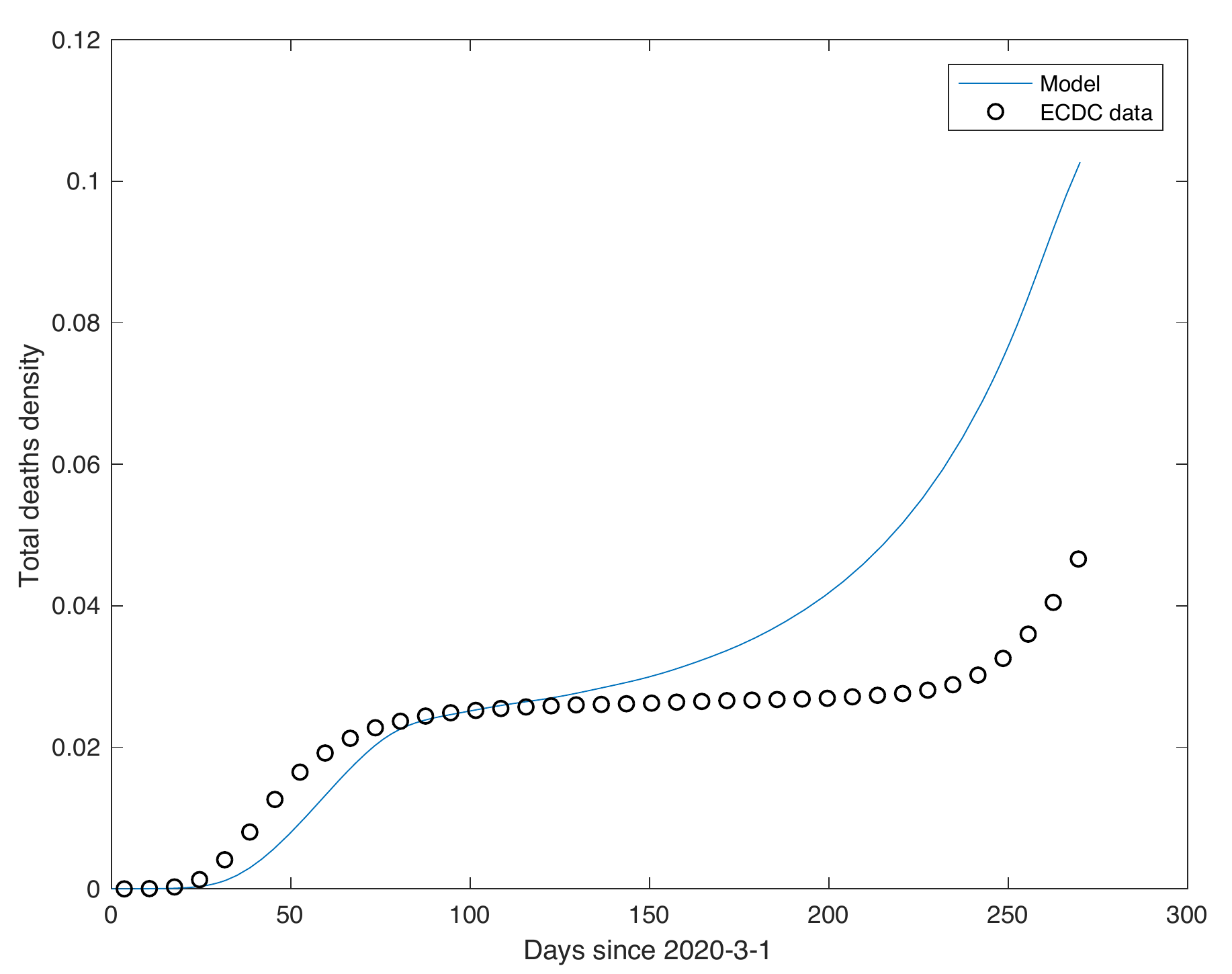}
      \caption{\textit{Comparison between the theoretical solution of our model for Deaths with real data provided by the database \textit{(Robert Koch Institut. Country data from Worldbank.org}, for Germany.}}
      \label{D_DE}
    \end{minipage}
  \hfill
\end{figure*}

\noindent Figs~(\ref{I_new_DE}) and (\ref{D_DE}) shows the comparison between the theoretical predictions for $\Delta I_{new}(t)$ and deaths and real data for Belgium (according to the database \textit{(Robert Koch Institut). Country data from Worldbank.org}). Notice in Fig.~\ref{I_new_DE} the prediction of the \textit{second wave of infection by SARS-CoV-2}
%%%%%%%%%%%%%%%%%%%%%%%%%%%%%%%%%%%%%%%%
%%%%%%%%%%%%%%%%%%%%%%%%%%%%%%%%%%%%%%%%
\section{Conclusions and Perspecives}\label{C}

We showed that our model is able to produce predictions not only on the first but also on the second or even the third waves of SARS-CoV2 infections. The theoretical predictions are in line with the official number of cases with minimal parameter fitting. We discussed the strengths and limitations of the proposed model regarding the long-term predictions and, above all, the duration of how long the lockdown and the quarantine measures should be taken in force in order to limit as much as possible the intensities of subsequent SARS-CoV-2 infection waves. This task has been carried out by taking into account the theoretical results recently appeared in literature \cite{sonnino} and without neglecting the delay in the reactions steps. Our model has been applied in two different situations: the spreading of the Coronavirus in a small Country (Belgium) and in big Countries (France and Germany). 

\noindent It is worth noting the \textit{degree of the flexibility} of our model. For example, let us suppose that we need to set up a model able to distinguish old population (over 65 year old) from the young one (with age not exceeding 35 years), by assuming that the older population is twice as likely to get infected by Coronavirus with respect to the younger one. In this case, it is just sufficient to replace the scheme $I+S\xrightarrow{\mu} 2I$ with the scheme
\begin{align}\label{C1}
&I+S_Y \xrightarrow{\mu_y} 2I\\
&I+2S_O \xrightarrow{\mu_o} 3I\nonumber\\
&S=S_Y+S_O\nonumber
\end{align}
\noindent with $S_Y$ and $S_o$ denoting the \textit{susceptible young people} and the \textit{susceptible old people}, respectively. Another example could be the following. Let us suppose that we need to distinguish two class of infected individuals: 

\noindent {\bf 1)} infected people (denoted by $I_1$) able to transmit the Coronavirus to susceptible according to the (standard) scheme $I_1+S\rightarrow 2I$;

\noindent {\bf 2)} Infected people (denoted by $I_2$) having the capacity to transmit the virus, say, 7 times higher with respect to the category {\bf 1)}. In this case, the corresponding scheme reads:
\begin{align}\label{C2}
&I_1+S \xrightarrow{\mu_1} 2I\\
&I_2+7S \xrightarrow{\mu_2} 8I\nonumber\\
&I=I_1+I_2\nonumber
\end{align}
\noindent It is then easy to write the ordinary differential equations associated to schemes (\ref {C1}) and (\ref {C2}). 

\noindent Let us now consider another aspect of the model. In the Subsection~(\ref{LQM}), we have introduced scheme~(\ref{LQM1}) that models the lockdown measures. As mentioned, such measures are imposed by national governments to all susceptible population. However, we can also take into consideration the hypothesis that these measures are not rigorously respected by the population and this for various reasons: neglect of the problem, depression due to prolonged isolation, lack of confidence in the measures adopted by the Government, desire to attend parties with friends and relatives, etc. Scheme~(\ref{LQM1}) still adapts to describe these kind of situations with the trick of replacing Fig.~\ref{LEP} with a curve that models the \textit{emotional behaviour} of susceptible people. The O.D.E.s read
\begin{align}\label{C3}
&{\dot S}=-\mu SI - k_E S(E_{Max}-S_E)+(1-k_E)(E_{Max}-E)\\
&{\dot S}_E=k_E SE-k^{-1}_ES_E\nonumber
\end{align}
\noindent where $E$ stands for \textit{Emotional}.

\noindent Finally, we mention that in ref.~\cite{sonnino2} we have incorporated real data into a stochastic model. The goal is to obtain a comparative analysis against the deterministic one, in order to use the new theoretical results to predict the number of new cases of infected people and to propose possible changes to the measures of isolation.

\end{document}